\documentclass[journal,comsoc]{IEEEtran}
\usepackage[T1]{fontenc}
\usepackage{enumerate}

\ifCLASSINFOpdf
  \usepackage[pdftex]{graphicx}

  \DeclareGraphicsExtensions{.pdf,.jpeg,.png}
\else

\fi

\usepackage{color,soul}

\usepackage{amsmath}

\usepackage[cmintegrals]{newtxmath}

\usepackage{algpseudocode}

\begin{document}

\title{Massive MIMO Adaptive Modulation and Coding Using Online Deep Learning Algorithm}

\author{Evgeny~Bobrov,
        Dmitry~Kropotov,
        Hao~Lu,
        and~Danila~Zaev

\thanks{The work has been supported by Huawei Technologies and Interdisciplinary Scientific and Educational School of Moscow University «Brain, Cognitive Systems, Artificial Intelligence».  Evgeny~Bobrov is with Moscow Research Center, Huawei Technologies, Russia, and M. V. Lomonosov Moscow State University, Russia (e-mail: eugenbobrov@ya.ru). Dmitry~Kropotov is with National Research University Higher School of Economics, Russia, and M. V. Lomonosov Moscow State University, Russia (e-mail: dkropotov@yandex.ru). Hao~Lu and Danila~Zaev are with Moscow Research Center, Huawei Technologies, Russia (e-mail: luhao12@huawei.com, zaev.da@gmail.com).}}

\maketitle

\begin{abstract}
The paper describes an online deep learning algorithm (ODL) for adaptive modulation and coding in massive MIMO. The algorithm is based on a fully connected neural network, which is initially trained on the output of the traditional algorithm and then incrementally retrained by the service feedback of its output. We show the advantage of our solution over the state-of-the-art Q-learning approach. We provide system-level simulation results to support this conclusion in various scenarios with different channel characteristics and different user speeds. Compared with traditional OLLA, the algorithm shows a 10\% to 20\% improvement in user throughput in the full-buffer case. 
\end{abstract}


\section{Introduction}

\IEEEPARstart{T}{he} adaptive modulation and coding (AMC) process carried out in the link adaptation is a crucial part of current wireless communication systems. It becomes especially important and challenging in massive MIMO systems with dynamic beamforming. Advanced AMC techniques allow a significant increase in the data rate that can be reliably transmitted \cite{chung2001degrees}.

Following New Radio (5G) downlink AMC procedure \cite{wannstrom2013lte}, user equipment (UE) has to suggest to the serving base station (BS) an appropriate modulation and coding scheme (MCS) to be used in the next transmission. The proposed MCS is provided by UE using a channel quality indicator (CQI). However, this indication is not enough for high-performance service. The first reason is that each CQI is associated with an interval of signal-to-inference-and-noise ratio (SINR), which could correspond to more than one MCS. In addition, in massive MIMO systems, the accuracy of CQI is limited by the number of specific antenna ports, which is usually less than the number of transmit antennas at BS. Due to this, BS cannot rely solely on the user's CQI report in MCS selection. That is why various AMC methods are proposed for this goal.

The well-known outer loop link adaptation (OLLA) technique was first proposed in \cite{sampath1997setting}. OLLA modifies the SINR CQI-based estimation by an offset \cite{song2013performance, pedersen2007frequency} which can be positive (making the MCS selection more conservative) or negative (when the CQI selection was too optimistic). This offset is updated based on transport blocks' transmission success rate so that the average block error rate is kept as close as possible to the predefined target \cite{blanquez2016eolla}.

It should be noted that the OLLA family of algorithms uses only the last binary acknowledgment information and does not take into account more refined SINR channel data, e.g., sounding (SRS) based measurements. Contrary to that, we offer an adaptive and self-learning method that predicts the next MCS using the available SINR-related measurements. The method performs both the mapping from SINR and channel data to the optimal MCS and the training (self-learning) in an online manner.

The main advantage of the proposed online deep learning (ODL) algorithm is its ability to adapt to different environments, different channel types, and different scenario conditions that BS cannot measure directly, e.g., UE speed. 
Due to the channel aging effect, user speed is an important hidden factor for the optimal choice of MCS, and it is hard to catch it with an offline pre-trained AI-based model. In the proposed approach, the model is able to adaptively learn the behavior of the UE and implicitly take into account its speed. In the state of the art, this challenge is called \b{concept drift}~\cite{gama2014survey}. It is described as a situation when some hidden features are important and change over time, but cannot be measured. This way, our task falls into the class of incremental learning \cite{krawczyk2018online} algorithms, which proceed with optimization in non-stationary environments such as the massive MIMO service of a mobile UE. The deep learning approach in massive MIMO scenarios was also studied in the work \cite{wen2018deep}.

Traditional OLLA adapts its offset based on HARQ acknowledgment (ACK/NACK) feedback for a transmitted transport block. The adaptation is done only if the transmission is performed. In this respect, the OLLA technique is highly dependent on traffic characteristics. If traffic is sparse compared with the channel variation, the OLLA adaptation may not achieve satisfactory quality. However, other modern techniques, like e.g., eOLLA \cite{blanquez2016eolla}, can update their offset independently of whether a transmission is carried out or not, which is very convenient for bursty traffic scenarios. In this manuscript, the proposed solution updates its parameters only using ACK/NACK feedback and assumes continuous (i.e. full-buffer) traffic. The proposed solution is fully compatible with 5G NR specifications (Release 15 or higher). It does not require any modification to the standard.

The novelty of our work is in the proposed scheme of online deep learning with a new optimization target. On the one hand, it is simpler and more effective than the existing Q-learning approach (\cite{mota2019adaptive, zhang2019deep}) to the AMC problem. On the one hand, it outperforms the basic OLLA approach because of the better utilization of the available channel//SINR information.

In this manuscript, machine training and execution are carried out exclusively on the base station side. The input (set of 'features') of the algorithm consists of the subband SINR measurements, CQI, time period from the last sounding, and the last reference signal received power (RSRP) measurement. Training data ('features' and 'labels') is collected in real-time and stored in a limited memory buffer. The computational complexity and storage requirements of the ODL approach have been investigated. Simulation results prove the stable behavior of the proposal and its uniform advantage over OLLA and Q-learning baselines. Quantitatively, the proposal increases throughput value compared to OLLA by 10\% to 20\%, depending on the agent speed.

We summarize the advantages of our proposal as follows: (i)~the ODL can adapt to different agent speeds, (ii)~the proposed approach is fully compliant with the existing NR 5G specifications, (iii)~the entire online machine learning process is conducted at the base station side, has feasible storage and computational overheads.

This paper is organized as follows. Section 2 briefly describes the massive MIMO model. Section 3 carries out the proposed algorithm structure, the neural network model, and the complexity of the online training with the sample buffer approach. Section 4 describes the simulation results. Section 5 contains the conclusion.

\section{System Model}
In the MIMO system, it is possible to send several information symbols to a multi-antenna user on a single physical resource. The number of such symbols is called the rank of the user. Under certain channel conditions, the higher rank can significantly increase the amount of transmitted information, but at the same time, it increases the requirements for channel quality. The single-user MIMO model is described by the following linear system:
 \begin{equation}\label{System Model}
  r = G (HWx+n).
 \end{equation}


Where $r \in \mathbb{C}^L$ is a vector of detected symbols at receiver, $x \in \mathbb{C}^L$ is a vector of sent symbols, $H \in \mathbb{C}^{R \times T}$ is a channel matrix, $W \in \mathbb{C}^{T \times L}$ is a precoding matrix, $G \in \mathbb{C}^{L \times R}$ is a detection matrix, and $n \sim \mathcal{CN}(0,I_L)$ is a noise-vector. The constant $T$ is the number of transmit antennas, $R$ is the number of receiver antennas, $L$ is the user rank. We assume they are related as follows: $L \leqslant R \leqslant T$. As for detection matrix $G$ we assume linear MMSE \cite{wubben2004near} and for the precoding $W$ we assume the SVD-based transmission scheme \cite{sun2010eigen}. 

The optimization objective is to maximize the expected throughput and was also considered, e.g., in \cite{fan2011understanding}.
The parameters of the model, including the bandwidth and the sounding period, are provided in the Table \ref{tab:configuration}.


\section{Structure of the proposed algorithm}

The general structure of the solution follows \cite{blanquez2016eolla}. The algorithm predicts the success acknowledgment (ACK) probability for each MCS given the available SINR measurements. 

We propose to consider the product of spectral efficiency and the probability of successful transmission, and maximize the resulting value over possible choices of MCS:
\begin{equation}\label{Proposed Method}
    \widehat{mcs}_{SE}(sinr) = \arg \max\limits_{mcs}  \big\{ p_w(ack|mcs, sinr)SE(mcs)\big\}
\end{equation}
This approach corresponds to the maximization of the expected throughput under the assumption of the Bernoulli probabilistic scheme.

Here, $p_w (ack|mcs, sinr)$ is a neural network model that predicts probabilities and has weights $w$ as the parameters for optimization. At the inference stage, the neural network takes as input the frequency-specific SINR estimations and an MCS and provides an acknowledgment probability as an output. The algorithm iterates through the MCS values and selects the scheme that provides the maximum expected throughput.

In the current state-of-the-art, there is a tendency to use the Q-learning (also called reinforcement learning) technique for the AMC problem \cite{mota2019adaptive, zhang2019deep}. This technique considers MCS selection as an agent action. While deep Q-learning is widely applied in wireless communication systems and can be applied to this task as well, we argue that this application is not natural. We propose an alternative scheme \eqref{Proposed Method} using classical deep learning that appears to be superior to Q-learning. 
Our choice of architecture is based on the following observations:

\begin{enumerate}[(a)]

    \item All actions in AMC are performed immediately and the reward delay is strictly specified in advance. The reward does not depend on the future actions, as, e.g., in a chess game that is modeled by Q-learning.

    \item There is no influence on the system from our actions. The actual SINR of the transmission is independent of the MCS we choose.

    \item The actual channel, the BS measurements, and the precoding are time-varying in general. Thus, we have access to the input data (features) and training outputs (labels) sequentially. Older data samples tend to become irrelevant.

\end{enumerate}

Observations (a) and (b) motivate the use of the traditional deep learning approach rather than Q-learning. We consider acknowledgment prediction as a binary classification problem and use the scheme \eqref{Proposed Method} to select the optimal MCS. Observation c) motivates the use of the online approach.

Compared with Q-learning, the main difference in our ODL approach is the use of a binary logarithmic loss function (log-loss) instead of Q-learning Temporal-Difference (TD)-Loss \cite{tesauro1995temporal}. This way, we move to the binary classification problem instead of maximizing the delayed rewards (a) and modeling the influence on the system of our actions (b).

Note that we do not need to model a chain of future actions for this type of optimization. Indeed, the proposed ODL method predicts only the MCS for the next transmission, while the Q-learning approach predicts a chain of future actions (which are enclosed in Q-values). Thus, the ODL method is more suitable for the MCS selection problem and, as we show later, provides more stable performance.

\begin{figure}

\centering

\includegraphics[width=\linewidth]{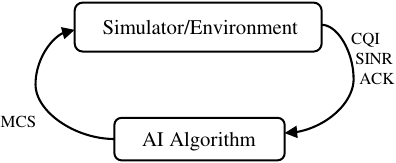}




\caption{Online Deep Learning algorithm block scheme.}

\label{fig:scheme}

\end{figure}

As a competitor to our solution, we consider the following Q-learning regression model \cite{mota2019adaptive, zhang2019deep}, which selects MCS based on the following maximization principle:

\begin{equation}\label{Alternative Method}
    \widetilde{mcs}_{SE}(sinr) = \arg \max\limits_{mcs}  \big\{q_w^{SE} (ack|mcs, sinr)\big\}
\end{equation}

Here, $q_w^{SE} (ack|mcs, sinr)$ is the neural network regression that predicts real scalar values. The Q-learning model is trained on the rewards $r(ack, mcs)=SE(mcs)[ack]$, where $[x]$ is the indicator function that returns $1$ if condition $x$ is true and $0$ otherwise. The condition $ack$ corresponds to the receipt of the success acknowledgment. We will discuss this in more detail in the next section.

\subsection{Neural Network Model}

In this work, we propose using the simplest neural network for binary classification without hidden layers (logistic regression). This model is lightweight, fast trainable, and robust to the environmental changes in the online-learning setting.

Our classification model uses the standard sigmoid function, which takes any real input $t$, and outputs a value between zero and one. The sigmoid function $\sigma : \mathbb{R} \rightarrow (0, 1)$ is defined as follows: $\sigma(t)=1/(1+e^{-t})$.

Thus, we can express the probability of receiving \textit{acknowledgement} in terms of the sigmoid function $\sigma$ depending on $mcs$ and $sinr$ arguments through the function $f_w$, which is the neural network function with weights $w$:
\begin{equation}\label{Sigmoid Function}
    p_w (ack|mcs, sinr) = \sigma (f_w (mcs, sinr)) ).
\end{equation}
The output of the model for a given vector of input features can be interpreted as a probability and serves as the basis for classification. The optimization method computes the log-loss for all the observations $n~\in~\{1~\dots~N\}$ on which it is trained. The function $J$ counts the log-probabilities of ACKs in the following way:
\begin{multline}\label{Classification Function}
   J(w)= \frac{1}{N} \sum_{n=1}^N \big(ack_n  \log  p_w (ack_n |sinr_n, mcs_n ) + \\ (1 - ack_n ) \log (1 - p_w (ack_n |sinr_n, mcs_n))\big) \rightarrow \max\limits_w
\end{multline}
Here $ack_n \in \{0, 1\}$ is the "true" \textit{acknowledgement}, which we get to know after the action is completed, and $p_w (ack_n|sinr_n,mcs_n)$ is the probability model of the $ack_n$ reception, which is a function of features: $\{sinr_n,mcs_n\}$.

For the Q-learning approach, we apply the MSE-Loss function to the reward. Since we do not have a delayed reward, this is the TD-Loss with $\gamma=0$ \cite{tesauro1995temporal}:
\begin{equation}\label{Regression Function}
    F(w)=\frac{1}{N} \sum_{n=1}^N \big(q_w (ack_n|sinr_n,mcs_n) -  r(ack_n,mcs_n )\big)^2 \rightarrow \min\limits_w
\end{equation}
\subsection{Proposed algorithm complexity}

For the process of online learning, we use the Adam \cite{kingma2014adam} method as one of the simplest gradient-based algorithms. It is worth noting that the previously obtained solution $w_t^*$ (optimal weights of the model) can be used as the starting point for the next re-training step $w_{t+1}^o$, resulting in: $w_{t+1}^o=w_t^*$. Therefore, in practical implementation, it is enough to just make a few gradient steps at the re-training step. 

\begin{figure}

\centering

\includegraphics[width=\linewidth]{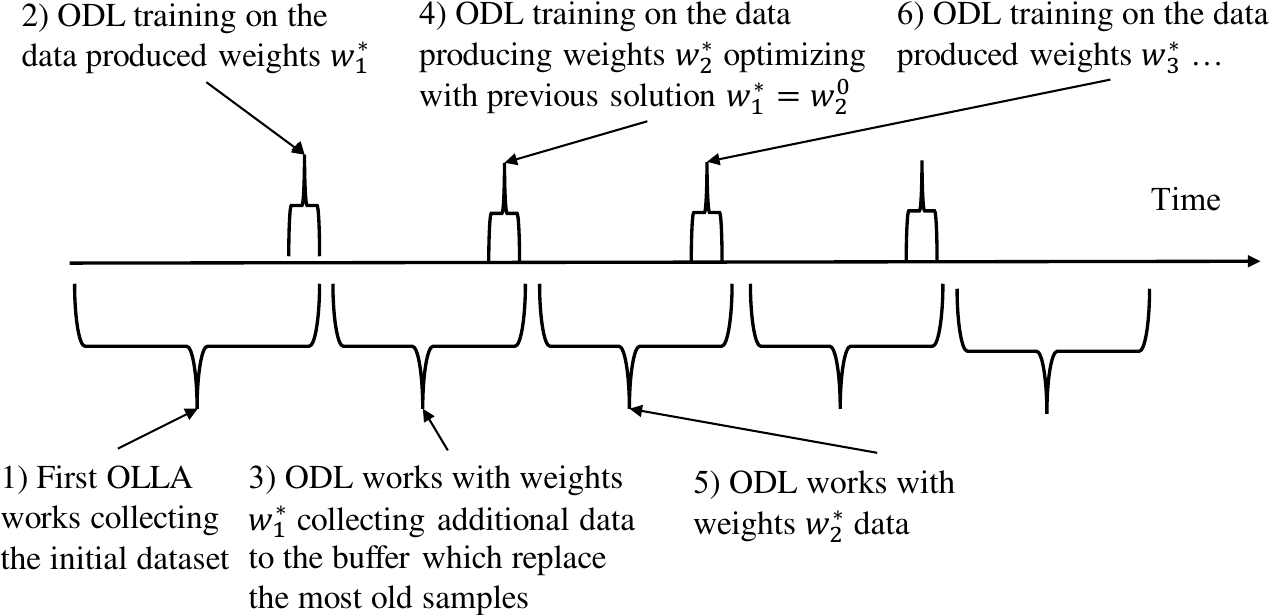}

\caption{Working Algorithm Time Axis.}

\label{fig:time}

\end{figure}

\begin{figure}
\centering
\includegraphics[width=\linewidth]{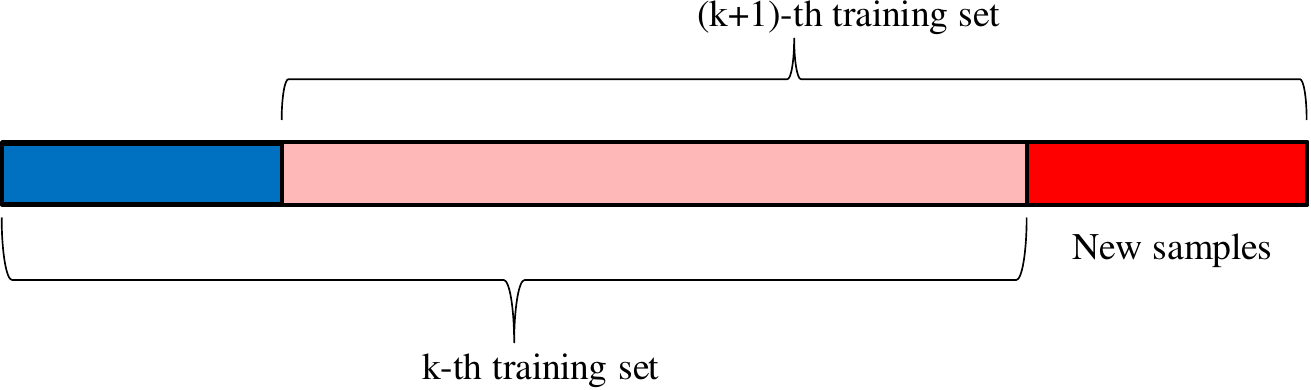}
\caption{Algorithm Sample Buffer.}
\label{fig:buffer}
\end{figure}

Since the algorithm works online, it needs to be retrained on the new data. We suggest using a buffer for every user containing the recent samples (transmission examples): features, the selected MCS, and the result of the transmission (ack/nack). Buffer samples are updated in FIFO order; the oldest samples are replaced with the newest ones. We can visualize this mechanism as follows: (Fig. \ref{fig:buffer}).

We propose adding new samples to the buffer with a (possibly adaptive) subsampling rate to avoid the situation where most features remain the same between the channel measurements. By doing so, we significantly reduce the memory buffer size and retraining speed without sacrificing prediction quality. The quality may even get better since we can expand the buffer to our storage limits. For our experiments with full-buffer users, we chose subsampling rate as an inverse probability of sounding length, excluding the pilot signals.

\begin{figure}

         \textbf{Parameters}: Initial value $y_o$ and step size $d$ of OLLA. Initial CQI $c_o$, target BLER $b$, buffer size $U$, retraining period:~$N$. 

         \textbf{Initialize}: OLLA: $y=y_o$, CQI: $c=c_o$, the sample buffer of the size $U$ and the neural network model $A(w)$ with number of nodes $P$ and number of connections $Q$.

         \textbf{Complexity}. Computations $\mathcal{O}(Q/N)$ and memory $\mathcal{O}(PU)$


\begin{algorithmic}

    \Procedure{An Agent Scheme}{}

\For{each of the first $U$ TTIs}
    \State Set $ MCS \gets \min(\max(\text{round}(c+y), 1), 29)$ 
    \State {Receive and put to the buffer the labels: \hspace*{15mm} ACK or  NACK: $a \in \{0, 1\}$, and the features: \newline \hspace*{15mm}
        CQI  $c\in  \{1 \dots n\}$, SINR $s \in \mathbb R^m$, \newline \hspace*{15mm} MCS $\in \{1 \dots 29\}$}.
    \State Update OLLA: $y \gets y + d a  - d (1 - a) (1 - b) / b$  \cite{blanquez2016eolla} 
    \State Train $A(w)$ targeting  $J$ \eqref{Classification Function} or $F$ \eqref{Regression Function} and using buffer
\EndFor

\For{each time frame $k = U + 1, U + 2 \dots $ }
     \State Set MCS $\gets A(w) $ NN prediction by \eqref{Proposed Method} or \eqref{Alternative Method}
     \State Receive and replace the oldest values of the buffer \hspace*{15mm}
        labels: ACK or NACK: $a \in \{0,1\}$, and features: \hspace*{15mm}
        CQI: $c \in \{1 \dots n\}$, SINR $s \in \mathbb R ^m $,  \newline \hspace*{15mm} MCS $\in  \{1 \dots 29\}$.

\If{$k \text{ mod } N = 0$}
    \State Initialize NN $w_k^o=w_{k-N}^*$ from the previous step
    \State Retrain $A(w)$ using \eqref{Classification Function} or \eqref{Regression Function} and memory buffer
    
\EndIf
    
\EndFor

    \EndProcedure

\end{algorithmic}

\caption{The scheme of both Online Deep Learning (proposed) and Q-learning with a sample buffer. The key difference between the algorithms is in the different activation and loss functions.}

\label{Scheme of the ODL}

\end{figure}

\subsection{The structure of the neural network}

\begin{figure}

    \centering

    \includegraphics[width=\linewidth]{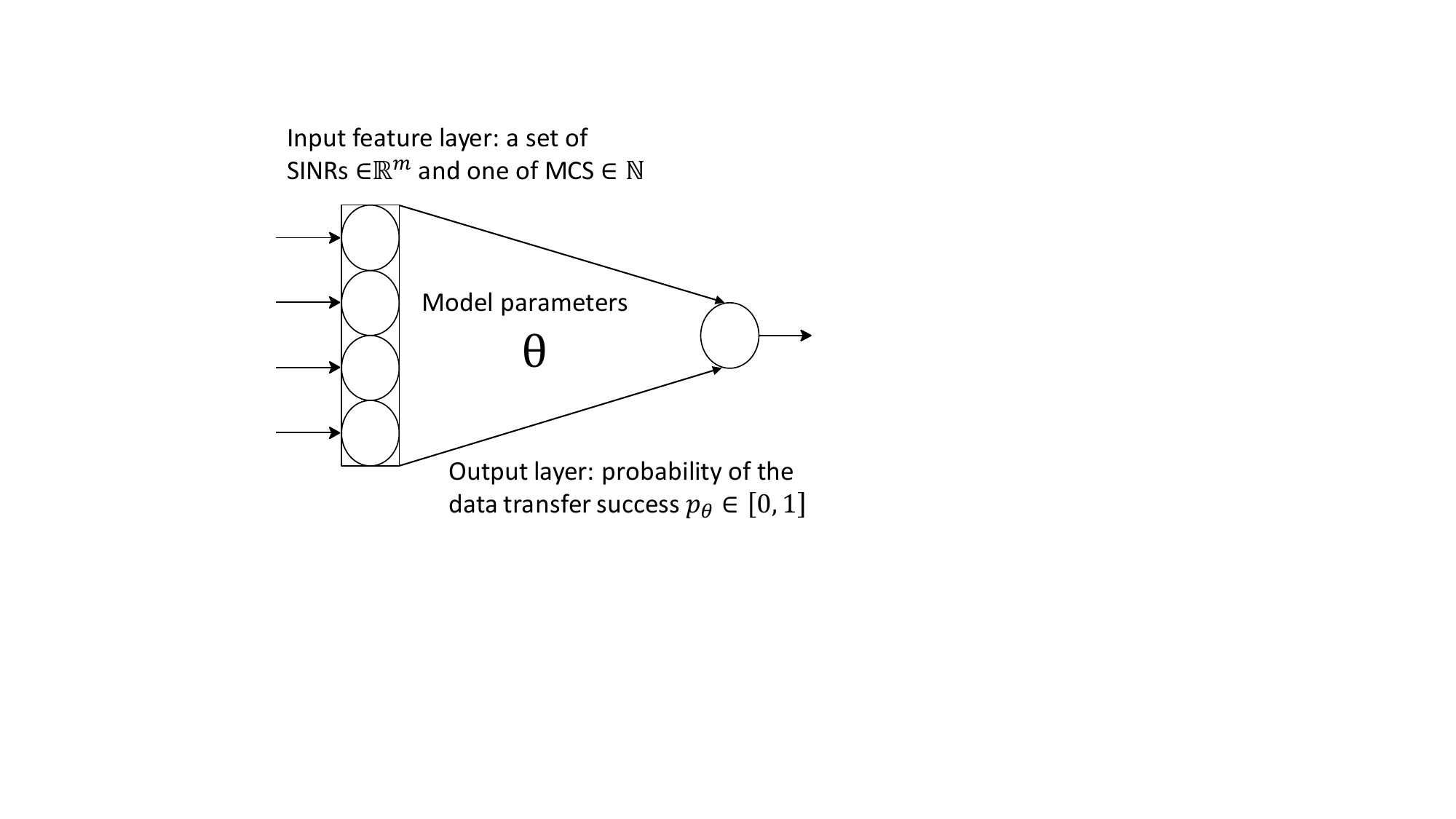}

    \caption{Block diagram of the neural network used.}

    \label{fig:nn_plot}

\end{figure}

The input of the neural network in the proposed approach and in the Q-learning approach is designed to be the same. The feature space of the algorithm consists of SINR for each user antenna ($RxNum=4$ for our experiments), reported CQI, the time interval from the last sounding, cell RSRP and one of the MCS values. Additional bias parameters are configured for each layer of the network. On the output layer the acknowledge success is predicted. We apply the standard scaling normalization method by subtracting the average value and dividing it by the standard deviation for each feature across all samples. The structure of the neural network is presented in Fig. \ref{fig:nn_plot}. 

The structure of ODL and Q-learning neural networks is the same in all aspects except the activation functions at the output layer. For the output layer, ODL uses the sigmoid function, while the Q-learning method uses the identity function. This difference is motivated by a difference in the problems the models solve. ODL solves the binary classification problem by predicting the probability of success, while the Q-learning model solves the regression problem by predicting real Q-values. Thus, we selected the activation functions that give the best quality results for each of the considered approaches.

\section{System-level simulation results}

First, we compare the proposed machine learning algorithm with the traditional OLLA method. The provided performance gains and losses are calculated with respect to OLLA performance. We have gotten stable, uniformly better results, which have never failed in our experiments. On average, the proposal increases throughput values from 12.64\% to 21.52\% depending on UE speed.

The advantage of the proposal is explained by the use of additional information based on SRS-based SINR measurements. We should also notice that the step-by-step behavior of OLLA is too conservative in a rapidly changing environment. 

\subsection{Quality improvement with machine learning}

We provide experimental results for different speeds, user ranks, and random seeds. Note that the proposed algorithm is not manually tuned for the various conditions: all its hyper-parameters remain the same. It is important since in the real-life commercial system, BS does not have information about the user speed and, especially, about the user environment (e.g., urban, rural, etc.).

\begin{figure}
\centering
\includegraphics[width=\linewidth]{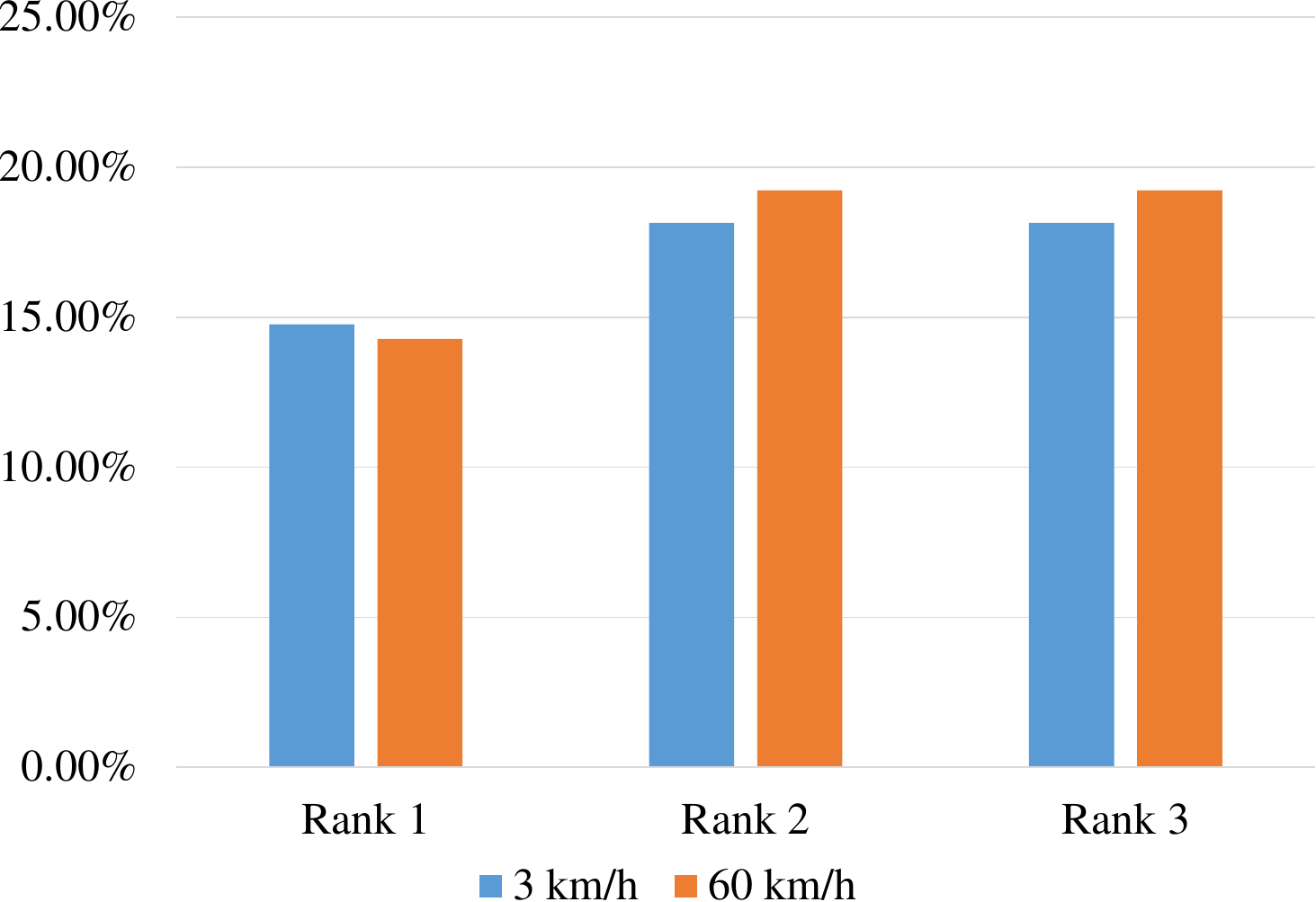}
\caption{The spectral Efficiency gain of ODL over OLLA. Average of 10 random seeds. Ranks 1, 2, and 3. Speeds 3km/h and 60 km/h.}
\label{fig:gains}
\end{figure}

The proposed online deep learning model performs the mapping from the SINR measurements to the optimal MCS. The most significant advantage is achieved on the rises and falls of the SINR quality because ODL is more adaptive to the instant SINR than OLLA and instantly converges to the optimal MCS. The following Fig. \ref{fig:througput} shows the uniform advantage of the ODL algorithm over OLLA.

\subsection{ODL and Q-learning performance comparison}\label{sec:Probabilistic Approach}

Next, we compare ODL performance with the performance of the Q-learning algorithm. Our simulation results show that the ODL method works uniformly better for all user ranks at a speed of 30 km/h and a random trajectory.

\begin{figure}
\centering
\includegraphics[width=\linewidth]{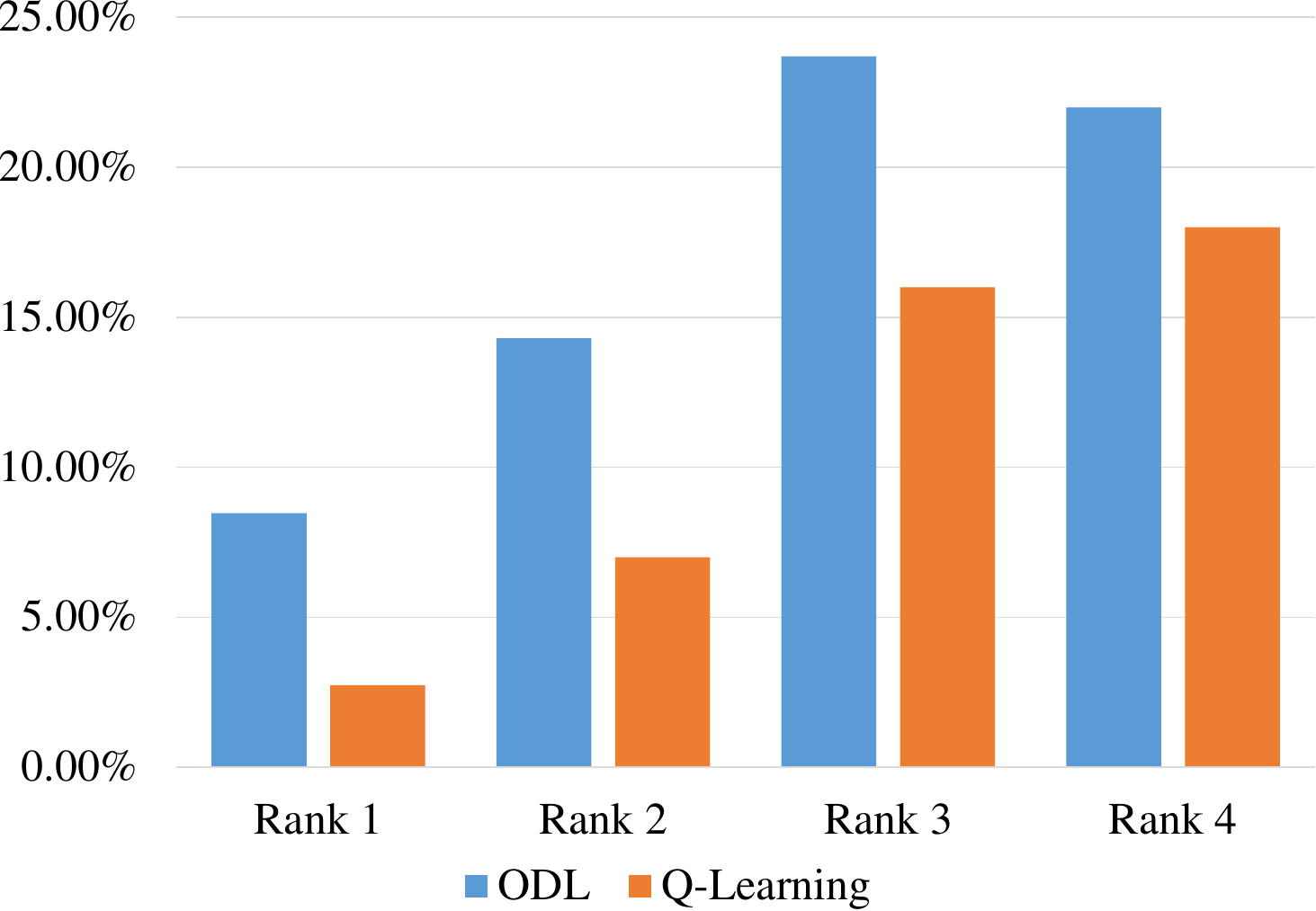}
\caption{Spectral Efficiency gain over OLLA of the two models: ODL and Q-learning. The agent speed is 30 km/h with a random trajectory for moving.}
\label{fig:gains_2}
\end{figure}

%
\begin{table}
\renewcommand{\arraystretch}{1.8}
\caption{System configuration of simulation-based experiments.}
\label{tab:configuration}
\centering
\begin{tabular}{|c|c|}
    \hline
    CellMaxPower & 40 dBm \\
    \hline
    ThermalNoisePower & -174 dBm/Hz \\
    \hline
    Bandwith & 20 MHz \\
    \hline
    TxAntNum, $T$ & 64 \\
    \hline
    RxAntNum, $R$ & 4 \\
    \hline
    Sounding Period & 5 ms \\
    \hline
\end{tabular}
\end{table}

\section{Conclusion}
This paper proposes a novel online deep learning solution for adaptive modulation and coding for massive MIMO systems. It  learns  to predict  the probability of transmission success for different MCS values and selects the MCS with the highest  expected  throughput. Simulation  results show that  the proposed  approach  outperforms  both the Q-learning  approach  and the traditional outer loop  link adaptation  method. When compared to standard OLLA, our method improves user performance by 10\% to 20\% in the full-buffer scenario. We provided an explanation for this advantage. The proposed approach has lower complexity than the Q-learning method and provides better and more stable performance. In addition, the proposed method is fully compatible with the current 5G RAN specifications. We hope that the analysis of the AMC problem provided in this paper will help to design better and simpler NN-based solutions for adaptive MCS selection in massive MIMO systems.

\begin{figure}
\centering
\includegraphics[width=\linewidth]{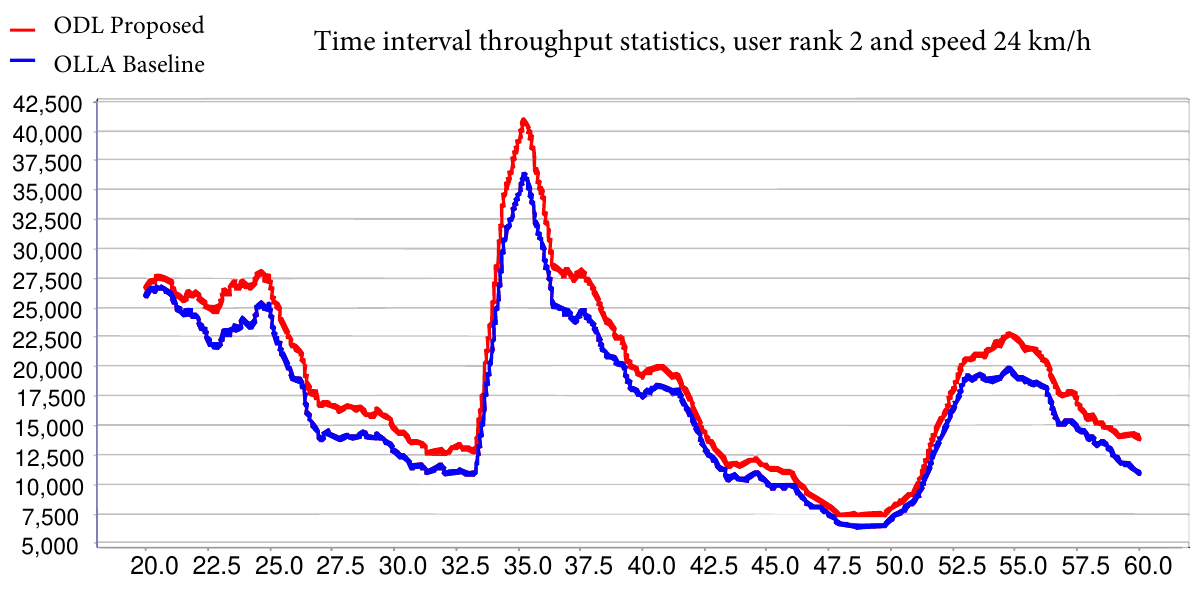}
\caption{Throughput statistics on a time interval, user rank 2 and 24 km/h speed. The red line is proposed ODL, the blue is OLLA.}
\label{fig:througput}
\end{figure}

\bibliographystyle{IEEEtran}
\bibliography{references}

\end{document}